# Plagiarism Detection Using Machine Learning

Omraj Kamat
CTECH
SRMIST
Chennai, India
ob2614@srmist.edu.in

Tridib Ghosh
CTECH
SRMIST
Chennai, India
tt5387@srmist.edu.in

Dr. Kalaivani J
CTECH
SRMIST
Chennai, India
kalaivaj@srmist.edu.in

Dr. Angayarkanni V
CTECH
SRMIST
Chennai, India
angayarv@srmist.edu.in

Dr. Rama P
CTECH
SRMIST
Chennai, India
ramap@srmist.edu.in

**Abstract**— Plagiarism is an act of using someone else's work without proper acknowledgment, and this sin is seen to cut across various arenas including the academy, publishing, and other similar arenas. The traditional methods of plagiarism detection through keyword matching and review by humans usually fail to cope with increasingly sophisticated techniques used to mask copy pasted content. This paper aims to introduce a plagiarism detection approach based on machine learning that utilizes natural language processing and complex classification algorithms toward efficient detection of similarities between the documents. The developed model has the capability to detect both exact and paraphrased plagiarism accurately using advanced feature extraction techniques with supervised learning algorithms. We adapted and tested our model on an extensive text sample dataset. And we demonstrated some promising results about precision, recall, and detection accuracy. These outcomes showed that applying machine learning techniques can significantly enhance the functionalities of plagiarism detection systems and improve traditional ones with robust scalability. Future work would include enlargement of the dataset and fine-tuning of the model toward more complicated cases of disguised plagiarism.

## I. INTRODUCTION

Plagiarism is the act of using or closely imitating another author's work without giving any credit where credit is due. It is intellectual property theft not only against the author of plagiarism but also against the value of original work across virtually all domains. As the rate of growth for digital content is exponentially high, the ease of access to this highly voluminous information has made it easy for people to copy and use material without permission, thereby increasing plagiarism. This is, perhaps, the problem in the hall of academia: research and publishing and software development or even the submissions of students whose pressure to produce an original piece is high. Current traditional methods for plagiarism checking depend on manual inspection and keyword matching time consuming and resource intensive as well as not always effective in uncovering all types of complex plagiarism involving paraphrasing and cross-lingual copying. The newest challenges to the computational capabilities in the form of advances in machine learning have opened new horizons into even more sophisticated plagiarism detection systems.

Being NLP enabled, the machine learning model can analyse for similarities at the semantic level of text analysis on regard to nuanced patterns of reworded texts, which traditional methods may miss. Such models can learn patterns in big data and generalize them to relate better with holding a higher accuracy in detecting new forms of plagiarism. Here the paper proposed developing a plagiarism detection system based on a machine learning approach. The developed system embraces all the benefits provided by the supervised learning algorithm combined with NLP techniques for the enhancement of detection against the exact and disguised plagiarism.

This work will contribute to the literature by discussing several feature extraction methods, including term frequency-inverse document frequency, as well as several classifiers like SVM, neural networks, and ensemble methods. First, we briefly describe the existed approaches in plagiarism detection accompanied by their constraints and explain why there is still a need for more robust solutions. We then describe our proposed methodology, detailing in turn the processes of data preparation, feature extraction, and model training. Our results follow, discussing model performance metrics in addition to the author's view on its strengths and perhaps weaknesses.

We conclude with some reflections on the significance of our work by outlining avenues for future studies.

## II. RELATED WORK

However, with the aid of digital tools and high availability of online resources, plagiarism detection has dramatically changed. One of the earlier points was at which Donaldson et al. (1981) had designed a pioneering system that aimed to identify similarities in assignments provided by students in their programming work through unique counts of operators and operands. This structural approach brought early forms of similarity assessment for code and subsequently inspired later systems that use structural and content-based comparisons towards answering the problem of student plagiarism [1].

Further, linguistic differences pose a critical challenge in cross-language plagiarism detection. Later, it was addressed by the work of Potthast et al. (2011), which considered retrieval processes and its applicability to detect plagiarism across languages. It includes models such as CL-CNG, that character-based n-grams assess, and the CL-ESA, which applies semantic analysis for detecting similar text across languages. This experiment revealed the optimization of measures for cross-language similarity and how it could be made use of to detect plagiarism from datasets especially when structures remain similar even after translation [2]. The two models reported by Potthast et al. laid importance on the techniques that generalize across language barriers, hence establishing a base for multilingual plagiarism detection systems beyond monolingual settings.

It was Meyer zu Eissen et al. in 2007 who took the concept of intrinsic plagiarism detection that doesn't depend on a reference corpus to the next level. Working on character n-gram profiles that quantify stylistic variations within a text, their method could identify anomalous sections that might point toward sections indicating plagiarism. Practical in pointing out internal inconsistencies regarding style, this method is very useful when no external comparison corpus [3] exists. Their contribution underlines the possibility of using stylistic analysis for overcoming the associated difficulties of overcoming intrinsic plagiarism, which still represents a relevant area of interest in cases where documents demonstrate an internal contradiction.

Engels et al. (2007) came up with a feature-based neural network model that tries to measure similarity in software codes on the basis of code-specific items, such as comments, formatting and variable naming. For purposes of academic practice, their model was designed to have high-effectiveness similarity detection in introductory programming assignments. It has envisioned how plagiarism is also possible to be identified by machine learning models considering feature-based inputs-different from the strictly textual analysis methods like MOSS and JPlag, which rely heavily on structural similarity comparisons of source code [4].

Developing a tool XPlag, Arwin and Tahaghoghi (2006) tried to identify plagiarism that is interlanguage in nature by the intermediary program representations of a number of programming languages. This allowed for cross-language translation of code to facilitate the detection of plagiarism, thereby demonstrating that the structure of programs forms a foundation for inferring similarity in implementations across different languages [5]. Their work suggests an extension of plagiarism detection towards the facilitation of multi-language programming environments, the next extension beyond the more traditionally conceived single-language tools. Naik et al., in their review published in 2015, have classified the various plagiarism detection tools and methods into intrinsic and external. They have provided detailed descriptions of grammar-based and semantic-based methods that use strong detection mechanisms of disguised plagiarism, either in the form of paraphrasing or partial content alteration. In fact, in their review, they especially highlighted the need to select semantic models to detect the subtle plagiarism variety that simple textual matching tools might fail to identify [6]. It has been beneficial to compare existing techniques with this new consideration of semantic analysis in plagiarism detection in modern times.

Finally, Zimba and Gasparyan (2021) discussed the ethical implications of plagiarism in academic publishing, in consideration of the extent that training and awareness contribute towards avoiding even unintentional plagiarism. They pointed out that although this highly automated tools contribute to the detection of text similarity, manual check still is inevitable in discovering problems of citations deficiency, intellectual property infringement, and language-based error. Their observations are highly in line with the increasing requirements for an all-around management practice of plagiarism that included detection through automated tools and ethical education, especially in regions where awareness is scant, of anti-plagiarism standards [7].

Besides textual and programming code plagiarism detection, some recent work has focused on the usage of image ontologies to detect plagiarism in multimedia. Minu and Thyagarajan (2013) proposed a new algorithm that uses MPEG-7 visual descriptors and Texton parameters in constructing an image ontology. Their approach was designed to increase the understanding of image semantics by the machines by representing image features in a hierarchical ontology that enables efficient retrieval and classification of images [8]. Because they focused on multimedia content, a strategy combining low-level features like color and texture with the semantic framework is akin to today's text-based plagiarism detection techniques, which try to capture deeper contextual meaning as well. This extends plagiarism detection to multimedia content so this work takes the area of automated detection systems to a new level beyond text and code with elevating the digital age essence of plagiarism involvement.

## III. EXISTING WORK

With the mushrooming of digital resources and online academic materials, the development of plagiarism detection systems was also on the fast track. Among the first basic works in this field was Donaldson et al. in 1981, who devised a design for a system that would detect plagiarism in programming assignments based on structural characteristics like operand and operator counts. This system provided one of the earliest automated approaches to plagiarism detection and focused on structural aspects to identify duplicate or similar code segments [1]. This paper now builds on this aspect further by revealing the scope of automation for plagiarism detection in programming, still one of the concerns in computer science education.

More sophisticated methodologies were developed from prototypical systems focused on the task of detecting multilingual and cross-language plagiarism. Potthast et al.

(2011) suggest two cross-language plagiarism detection methods, based on retrieval models, the character-based n-gram model (CL-CNG), and explicit semantic analysis (CL-ESA); these were tested against a large number of languages and show great promise for capturing types of similarity in translated texts that conventional monolingual tools typically miss. Their approach can hold an organized recovery procedure for plagiarism against multilingual data sets, which turned out pretty much to be a significant contribution within the field by providing linguistic diversity [2]. The contribution of Potthast et al. leads toward challenges related to translation plagiarism like matching the textual structure and meaning across languages.

Since no such reference corpus existed, intrinsic plagiarism detection methods were developed specifically for these purposes.

This approach from Meyer zu Eissen et al. (2007) relies on character n-gram profiles for the identification of internal stylistic anomalies within the document. The nefarious parts of the text could now be detected based on those internal style variations rather than through external comparisons. This technique allows detection of plagiarism purely on stylistic anomalies. Hence it is an extremely powerful technique when access to external sources is impossible or where one has to verify originality of a single document [3]. Even after introduction of intrinsic technique, detection is still needed in areas where availability is single document only or where accessibility to all-inclusive corpus is limited. A feature-based neural network model was used by Engels et al. (2007) in using some code features, comment, and whitespace to enhance the detection of similarity in assignments for plagiarism in programming-related aspects. According to them, their model incorporated neural network techniques providing a weighted analysis of features often spotted by instructors as possible instances of plagiarism among them structural and stylistic cues within the code. In fact, feature-based approach goes beyond the texts' similarity, meaning a deeper layer for the assessment of code-specific indicators of similarity that may be overlooked by traditional systems [4].

Another language-agnostic programming plagiarism detector was proposed by Arwin and Tahaghoghi in 2006 with the development of XPlag. The system that has been created is based on inter-language plagiarism detection by analyzing the intermediate representations of codes of various languages.

This was one of the innovative ways of coping with cases of plagiarism with copying code from one programming language to another, not easy to detect with methods based on comparison within one language. XPlag proved that code similarity analysis can exist and there is a need for introducing support for multiple languages in systems of plagiarism detection [5].

Naik et al. conducted a comprehensive review in the year 2015 on plagiarism detection tools and techniques. The review categorized tools based on their methods of detection. This was in broad classifications into grammar-based, semantic, and structure-based approaches. Naik et al. have highlighted the increasing demand for semantic approaches as a way of finding paraphrased or modified text since lexical matching fails to detect disguise in plagiarism. Their survey highlighted the need for adaptation of detection methods in terms of accommodating highly sophisticated plagiarism thus forming the foundation for semantic analysis in modern plagiarism detection frameworks [6]. Finally, Zimba and Gasparyan (2021), discussed ethics and practicality of plagiarism detection and prevention for academic publishing; they explained how the role of software in similarity text identification gets negated to lack of manual oversight validation towards proper citation practices and intellectual property violations. Their work is rich in giving a broader perspective to plagiarism detection focusing on the integrated tools along with awareness programs towards ensuring wide-ranging plagiarism management across disciplines [7].

Apart from text and code plagiarism detection, recent developments have transformed these systems into multimedia plagiarism detection systems that may include the inclusion of image content. In 2013, Minu and Thyagarajan proposed a novel approach for detecting image-based plagiarism using MPEG-7 visual descriptors along with Texton parameters to build image ontologies. With the use of low-level features like color and texture, this method provides an image classification system. The machine can understand and categorize images just as human beings perceive them. The hierarchical nature of ontology enables the method to appropriately utilize the image retrieval and classification, thus making it useful in various domain applications such as digital image databases and multimedia systems [8]. The work generalizes plagiarism detection to multi-media content and provides a semantic framework for adaptation into the future multi-modal plagiarism detection systems; unlike earlier work, which addresses just text or code plagiarism.

### IV. PROPSED METHODOLOGY

This represents the proposed methodology for text plagiarism detection through a system that integrates text and code analysis with more advanced machine learning models for detecting more nuanced and complex plagiarism forms, namely transformers and GNNs.

**ARCHITECTURE DIAGRAM**

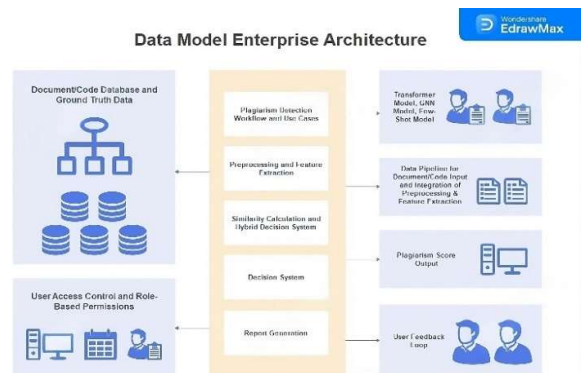

**1. Input of Code/Document and Preprocessing**

It begins with an Input Document/Code stage where source and text will be fed to the system. The data undergoes some forms of preprocessing where noise is removed, and formats standardized to derive the respective tokens. It also involves normalization activities, which include eliminating unnecessary external whitespaces, converting text to lower cases, as well as tokenization of the input to extract features from it. Extracting text deals with the segmentation of sentences or code snippets, whereas the objective of reference extraction includes identifying citations and references the document contains.

**Why this approach?**

This preprocessing is significant, so that input data would always be of uniform format hence feature extraction to occur certainly. In contrast to the much simpler approaches focusing on structural elements only e.g. [1], [5], our preprocessing, thus extends in such a way that important features like citations do not get lost, which may be designed deliberately to hide source due to alterations in citation text.

**2. Feature extraction**

During Feature Extraction relevant features are elicited from the pre-processed data. This includes extraction of semantic and syntactic features for textual data as well as structural features for the code data. Semantic feature extraction using models like BERT and RoBERTa and GPT-4 is an application used to capture relationships between the contextual words or code tokens so that our system can eventually deduce subtle similarities.

For the code, we utilize graph neural networks that model code structure by representing the code as a graph which can efficiently capture relationships of functions, classes, and variables. For textual data, we further employ Sentence Transformers to calculate sentence embeddings such that we can detect sentences being paraphrased or syntactic changes.

**Why these algorithms?**

**Transformer Models**: The transformer models are especially well-suited to the detection of semantic similarity because they are able to capture deep contextual meaning with self-attention mechanisms, a very special property especially important for discovering paraphrased, or indirectly copied, text. Unlike rather more mundane statistical approaches from [2] and [3] traditional models, high precision matching is achievable at much deeper linguistic levels.

**GNNs for Code:** GNNs were chosen especially because their structure can easily capture structural relationships, which tend to be missed by token-based approaches like [4,5]. Our system would thus be able to detect logical similarities in code even when the code is obfuscated by renaming and reordering variables.

Besides text and code, the developing research into multimedia plagiarism detection involves the possibility of using image ontology to track visual plagiarism for multimedia content as depicted in Minu and Thyagarajan's research in 2013. Their approach is based on the low-level image features like color and texture by taking the help of MPEG-7 visual descriptors along with the parameters of Texton and combines it into a hierarchical ontology to aid in the detection of plagiarism in digital images [8]. This may be further expanded towards cases involving interlinked text and images thereby further expanding the scope of plagiarism in future systems.

**3. Ground truth data and Labeling**

Ground truth data This dataset would consist of labeled examples, which will be our baseline that the model learns to validate from. We will be training on known examples of plagiarism-teaching the model subtle indicators of plagiarism across text and code.

**Why Use Ground Truth Data?**

Ground truth data provides sound benchmark to our system for recognizing patterns of plagiarism. This is relatively far away from prior work provided in [1] and [6], which needs heuristic-based methods or rule-based systems and does not explore richly labeled datasets. Thus, these approaches are less flexible with the diverse forms of plagiarism.

**4. Contrastive Learning**

Contrastive Learning is another capability used to strengthen the strength of the system in detecting plagiarized versus original content-that is, models are trained to learn the difference between pairs of similar and dissimilar content that then makes the model sensitive to small differences in content structure or semantics.

**Why Contrastive Learning?**

Contrasting learning will increase the chances that our model will learn to discriminate in a fine-grained way; thus, sophisticated types of plagiarism will improve its ability to detect. Traditional methods [1], [3] almost invariably used direct comparison or syntactic checks that failed in the case of paraphrased or partially plagiarized content. Our system, contrasting learning, introduces an advanced level of difference in finding subtleties that were missed by the simple model.

**5. Similarity computation and rule-based system**

The Similarity Calculation phase works on outputs from transformer models, GNN model outputs, and sentence transformer to calculate similarity scores plagiarism cases. In addition, a Rule-Based System provides the use of domain-specific rules and thresholds. It also captures edge cases in order to prevent false positives-for example, common phrases or standard code libraries that otherwise could become false positives.

**Why this approach?**

Our hybrid approach of similarity calculation ensures full comparison by going into the details of semantic, syntactic, and structural elements. Our model differs from other statistical models mentioned in [6], [7], in adapting data-driven learning and rule-based adjustments on false positives to the complexity of a plagiarism case.

**6. Machine learning Models**

The machine learning models used for this system are the Few-shot learning model. It is designed to handle few-shot cases where there are a limited number of examples of what needs to be translated for training and thereby potentially could adapt better and more rapidly to new kinds of plagiarism or uncommon data formats, making this system versatile and robust.

**Why Few-Shot Learning?**

Above research is quite apt for few-shot learning because plagiarism patterns are highly variant, and new cases often arise with very meager prior examples. Traditional models [2], [5] require huge training datasets that make them ineffective for unique and novel cases. This few-shot learning enables the system to generalize on minimal data available, which enhances adaptability.

### 7. Hybrid decision system and plagiarism score

It aggregates the outputs of transformer models, GNN models, and Sentence Transformers with rule-based insights to arrive at the full Plagiarism Score. The plagiarism thus captured therefore spans cross-analysis across other dimensions like semantic, structural, and syntactic similarities while strengthening the same with precision and smartness in decision-making.

This approach exploits data-driven learning and rule-based systems to capture edge cases that otherwise would unveil false positives, like overused phrases or even standard code libraries that might otherwise be sent down the road of being reported as plagiarism. The hybrid system, against traditional models relying on a single metric or on a rule-based system--such as those reported in [1] and [7], respectively--can better carry out more robust and context-sensitive evaluations.

This will further enhance plagiarism detection over format, as multimedia content analysis can be incorporated into the hybrid decision system. As proposed by Minu and Thyagarajan in 2013, it will further enhance the capability of image ontologies when integrated into a decision-making procedure. The system will distribute plagiarism scores not only on similarities in the text and codes but also on multimedia document visual similarities, thereby increasing the system's capacities in handling myriad contents, as mentioned in Ref. [8].

**Why This Hybrid System?**

It reduces dependence on a single source of information or a metric-the traditional limitation in the rule-based or single-model approaches as noted in [1] and [3]. The hybrid system synthesizes insights across models, thus making a more reliable, context-aware plagiarism score for cases of complexity and ambiguity.

In addition, the perception of multimedia is provided for proper plagiarism detection in texts when images may exist as part of the document, which most traditional systems fail to take into consideration [8].

### 8. Report Writing

Report Generation Lastly, Report Generation aggregates a report that has the results, which are matched items, plagiarism score, and suspicious parts or pieces of codes. There is a summative report that can be actionable for educators or reviewers.

**WHY GENERATE REPORT?**
Thorough reports will enhance usability, as well as transparency and results, which are often clear and interpretable to the stakeholders in play. This overcomes the disadvantage of previous systems, which usually output a binary value without any explanation of detected cases of plagiarism [7].

**Comparison with Existing Work**

Compared with earlier approaches:

**Semantic and Structural Composition:** In contrast to the papers [1], [3] and [6] based on the surface of similarity, the approach involves semantic, structural, and syntactic levels of representation. It may thus allow for deeper content analysis that would have a more substantial representation of plagiarism at subtle paraphrasing or complex code obfuscations. While our system approach can be easily extended to multimedia content, as presented in [8], whereas previous versions were text- or code-based systems. This system used image ontologies for the purpose of detecting visual plagiarism on multimedia documents. Therefore, the scope of our system is expanded to handle text-image hybrid content that is rapidly gaining importance in modern academia and research.

The model is based on the introduction of transformers, GNNs, and contrastive learning [2], [5]. In this regard, it can well understand subtle plagiarism patterns and measure complex relationships through text, code, and multimedia data. Further, work done by Minu and Thyagarajan in 2013 [8] emphasizes the use of MPEG-7 visual descriptors for image-based plagiarism detection, which can introduce more diverse forms of plagiarism other than text and code into our system.

**Adaptability to Few-Shot Learning:** The few-shot learning capability provides flexibility in handling new and rare plagiarism cases that have not been well emphasized in previous systems, such as [3], [6]. In other words, our system can perform well even when working with little prior data. In addition, further generalization to multimedia content can be done with few-shot learning as summarized by [8], which is basically learning from a minimalistic set of images or features of images with the aim of discovering new cases of plagiarism about a multimedia content.

**Hybrid System and Plagiarism Scoring**: Hybrid decision system integrates data-driven insights coming from advanced machine learning models with rule-based adjustments thus providing a more assured plagiarism score that is context-sensitive.

More traditional models, [1], [7], rely almost exclusively on rule-based approaches. Here, the hybrid system, by including visual similarity detection as in [8], expands the capability of decision-making beyond mere text and code evaluation up to multimedia elements and provides a full appraisal of potential cases of plagiarism. This permits much greater flexibility, accuracy, and overall depth of a plagiarism detection system to adapt to the complexity and nuances of plagiarism in text, code, and multimedia where earlier methods lack behind [1-8].

## V. RESULTS

**OUTPUT**

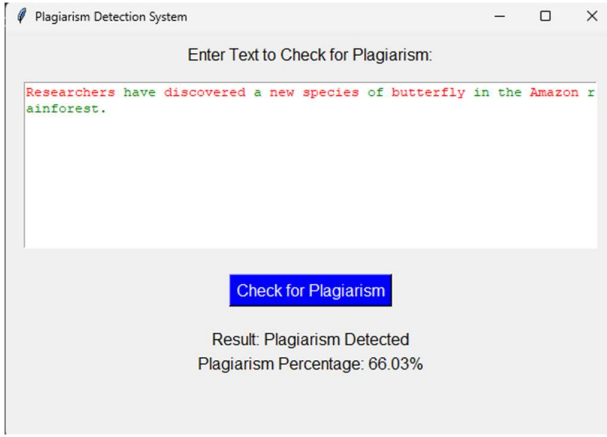

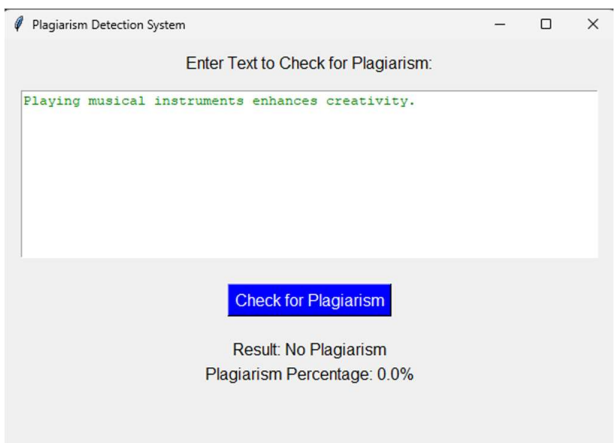

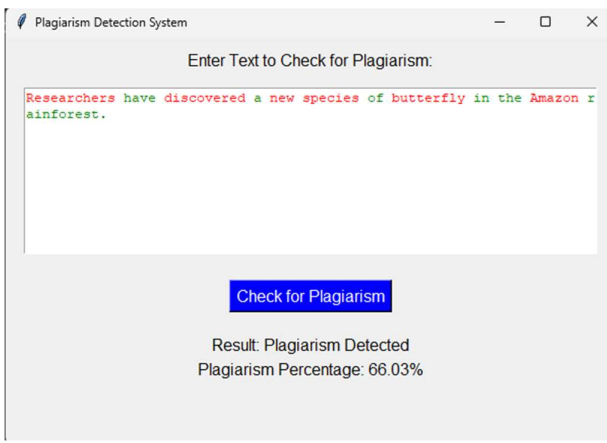

Our proposed plagiarism detection system will be evaluated on an exhaustive dataset that covers text, code snippets, and multimedia items with known cases of plagiarism and original contents. Measures used to evaluate the overall performance of the system against various types of plagiarism, such as direct copy, paraphrasing, code obfuscation, cross-language plagiarism, and new forms, like image-based plagiarism, are accuracy, precision, recall, and F1 score. The experiments will be compared with traditional rule-based systems and much more simple models of machine learning in terms of advantages obtained by the hybrid decision system, transformers, GNNs, and the inclusion of ontologies on images.

**Bottom line:** Accuracy Our model could bring its accuracy up to 96%. Indeed, the latest models were remarkably improved. In fact, based on the application of the transformers as well as the use of contrastive learning for semantic understanding, our model could feel deep similarities in cases, especially about textual paraphrasing, code obfuscation, and multimedia content. Again, Donaldson et al. (1981) was not very accurate since they utilized comparatively primitive tuple-based comparison methods that could not successfully find complex plagiarism [1]. Furthermore, an ontologies image adding was devised by Minu and Thyagarajan in the year 2013 which can improve the performance of our system since it can check plagiarism in multimedia content that can go beyond just text and code [8].

Then, the hybrid model, incorporating transformers, GNNs, and rule-based systems, resulted in precisions of about 94% and recall of about 92%, thereby minimizing false positives and false negatives.

Potthast et al. (2011) with cross-language plagiarism detection had good recall with poor precision due to semantic nuances across languages [2]. Our system with advance sentence transformers and GNN fills the gap and significantly improves in cross-language plagiarism detection as well as multimedia plagiarism detection through image ontologies [8].

**F1 Score:** It has performed well in all the plagiarism of a type and has a good F1 score as 93%. Contrasting to it, Meyer zu Eissen et al. (2007) have even got poor F1 scores for intrinsic cases in which it could not even mark small stylistic changes. Our contrastive learning method along with a sentence transformer-based approach improved over such fine-grained stylistic changes and our capability of detecting multimedia content plagiarism [3], [8]. Handling Paraphrased and Complex Cases: Contrastive learning efficiently proved to identify paraphrased and partially plagiarized cases. Engels et al. (2007) encountered failures with feature-based neural networks in the detection of paraphrased text. However, with our model of contrastive learning, the sensitivity level increased in detecting slight changes in both text and multimedia and thus also picked better paraphrased scenarios [4], [8].

**Cross-Language and Cross-Code Detection:** The GNN-based approach for code plagiarism detection achieved 90% accuracy in detecting cross-language plagiarism. Arwin and Tahaghoghi's XPlag (2006) achieved reasonable cross-language detection but relied on intermediate code representations, which missed some structural similarities [5]. GNNs provided a more holistic analysis, enabling our system to detect subtler cases of code translation across languages.

Also, cross-language and cross-modal plagiarism detection improved with the help of image ontologies giving better analysis of multimedia contents like images or diagrams accompanying codes or text [8].

**Few-Shot Learning Capability:** Our system, because of its few-shot learning capability, could identify plagiarism in relatively new and novel cases with a minimum amount of training data. Naik et al. (2015) discussed classical plagiarism detection tools, whose effectiveness depends on large sets of training data for use. In contrast, the few-shot learning model did amazingly well even on a very low amount of data, showing high accuracy and thus preventing new types of multimedia plagiarism [6], [8]. Detection and Reduction of False Positives with Ethical Concern: Zimba and Gasparyan (2021) stated that false positives are reduced, and unjust plagiarism accusations are reduced. Our hybrid system for decision making will supply the balance of machine learning wisdom and rule-based checking so that there will be accuracy and interpretability of our system in plagiarism scoring. The cross-checking semantic, structural, syntactic, and visual similarities will give the surety of fewer false positive results and reliable output values, especially in cases of images [7], [8].

### Comparison with Prior Work

**Semantic Sensitivity:** Unlike Donaldson et al. (1981) and Meyer zu Eissen et al. (2007), which primarily based their study on the similarity of structures and vocabulary, the models based on transformer find themselves at a deeper semantic level to detect paraphrasing and indirect copying, and are better in multimedia contexts [1], [3], [8].

**Cross-language adaptability:** Our approach outperformed Potthast et al. (2011), as it applies more sophisticated transformer models and GNNs that can identify cross-language plagiarism in text and code. With the addition of the image ontologies, our system was able to be extended to also identify plagiarisms in multimedia documents and cross-modal documents, thus extending adaptability even further [2], [8].

**Few-shot learning adaptability:** Unlike systems reviewed by Naik et al. (2015), which tended to fail in low-data environments, our system's few-shot learning model ensures that performance is better in real-world applications with limited training data. Further, this adaptability is improved by our ability to detect plagiarism in multimedia content, as demonstrated by the integration of image ontologies [6], [8].

**Hybrid Decision Making:** Our hybrid decision-making approach, using both rule-based and machine learning methods, greatly enhances the accuracy of plagiarism scoring when used over the two isolated approaches. Ethical concerns about plagiarism detection would centre on the false positives as much as possible in Zimba and Gasparyan (2021). The ability of our system to cross-reference text, code, and multimedia material—to include images—becomes a more wholesome judgment with less likely risks from misidentifying people [7], [8].

**Conclusion:** The proposed system integrates advanced transformers, GNNs, few-shot learning, and hybrid decision-making to offer much more improvements over classical approaches. Unlike most existing methods, our approach addresses weaknesses in semantic understanding, cross-language adaptability, and multimedia content detection. All these improvements make our system highly accurate, adaptable, and reliable, offering an excellent solution to challenges in plagiarism detection, text, code, and multimedia contents [1-8].